\documentclass{article}

\usepackage{arxiv}

\usepackage[utf8]{inputenc} 
\usepackage[T1]{fontenc}    
\usepackage{hyperref}       
\usepackage{url}            
\usepackage{booktabs}       
\usepackage{amsfonts}       
\usepackage{nicefrac}       
\usepackage{microtype}      
\usepackage{graphicx}
\usepackage[super,sort&compress,comma]{natbib} 
\usepackage{doi}
\usepackage{float}

\usepackage{xcolor}
\newcommand{\pg}[1]{{ #1 }}

\usepackage{gensymb}
\usepackage{amssymb}

\title{How do interfaces alter the dynamics of supercooled water?}

\author{ \href{https://orcid.org/0000-0001-7671-4825}{\includegraphics[scale=0.06]{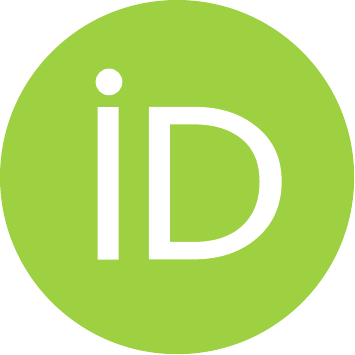}\hspace{1mm}Piero~Gasparotto}\thanks{Corresponding author.} \\
	Scientific Computing Division\\
	Paul Scherrer Institute\\
	Villigen 5232, Switzerland\\
	\texttt{piero.gasparotto@gmail.com} \\
	\And
	\href{https://orcid.org/0000-0001-6790-4301}{\includegraphics[scale=0.06]{orcid.pdf}\hspace{1mm}Martin~Fitzner} \\
	Thomas Young Centre \\
	London Centre for Nanotechnology and \\ 
	Department of Physics and Astronomy \\
	University College London \\
	London WC1E 6BT, United Kingdom \\
	\And
	\href{https://orcid.org/0000-0003-2708-8711}{\includegraphics[scale=0.06]{orcid.pdf}\hspace{1mm}Stephen J.~Cox} \\
	Yusuf Hamied Department of Chemistry  \\ 
	University of Cambridge \\
	Cambridge CB2 1EW, United Kingdom \\
	\And
	\href{https://orcid.org/0000-0002-6156-7399}{\includegraphics[scale=0.06]{orcid.pdf}\hspace{1mm}Gabriele C.~Sosso} \\
	Department of Chemistry \\
	University of Warwick \\
	Coventry CV4 7AL, United Kingdom \\
	\And
	\href{https://orcid.org/0000-0002-9169-169X}{\includegraphics[scale=0.06]{orcid.pdf}\hspace{1mm}Angelos~Michaelides}\thanks{Corresponding author.} \\
	Yusuf Hamied Department of Chemistry  \\ 
	University of Cambridge \\
	Cambridge CB2 1EW, United Kingdom \\
	\texttt{am452@cam.ac.uk} \\
}

\renewcommand{\headeright}{10.1039/D2NR00387B}
\renewcommand{\undertitle}{10.1039/D2NR00387B}
\renewcommand{\shorttitle}{How do interfaces alter the dynamics of supercooled water?}

\hypersetup{
pdftitle={A template for the arxiv style},
pdfsubject={q-bio.NC, q-bio.QM},
pdfauthor={David S.~Hippocampus, Elias D.~Striatum},
pdfkeywords={First keyword, Second keyword, More},
}

\begin{document}
\maketitle

\begin{abstract}
The structure of liquid water in the proximity of an interface can deviate significantly from that of bulk water, with surface-induced structural perturbations typically converging to bulk values at about $\sim$1\,nm from the interface. 
While these structural changes are well established 
it is, in contrast, less clear how an interface perturbs the dynamics of water molecules within the liquid. 
Here, through an extensive set of molecular dynamics simulations of supercooled bulk and interfacial water films and nano-droplets, we 
observe the formation of persistent, spatially extended dynamical domains in which the average mobility varies as a function of the distance from the interface. 
This is in stark contrast with the dynamical heterogeneity observed in bulk water, where these domains average out spatially over time. 
We also find that the dynamical response of water to an interface depends critically on the nature of the interface and on the choice of interface definition. 
Overall these results reveal a richness in the dynamics of interfacial water that opens up the prospect of tuning the dynamical response of water through specific modifications of the interface structure or confining material.
\end{abstract}


\section{Introduction}

In apparent contradiction with everyday experience, it is surprisingly difficult to crystallize water into ice. Liquid water can remain in a supercooled state at below 0\,\degree C for hours and even days. In the absence of impurities, the homogeneous freezing point of water is $-41$\,\degree C and experimental evidence shows that nanodroplets, composed of a few thousand water molecules, can be cooled down to $-70$\,\degree C without a hint of crystallization~\cite{jeffery1997homogeneous,sellberg2014ultrafast,laksmono2015anomalous,kim2017maxima}. The study of supercooled liquid water dates back to at least the time of Fahrenheit and his seminal temperature-defining measurements~\cite{fahrenheit1724viii}. Supercooled liquid water droplets are ubiquitous in clouds, while plants and mammals adapted for survival in cold climates exploit supercooled water for survival~\cite{valerio1992fish}. The study of supercooled water is also critical for rationalising the anomalous properties of water and as a means to understand ice formation~\cite{moore2011structural}.

Previous studies have shown that supercooled liquids (particularly glass forming liquids) exhibit very interesting dynamical properties. Specifically, upon supercooling, a phenomenon known as dynamical heterogeneity (DH) emerges~\cite{kawasaki2007correlation,la2004static,handle2018adam,ediger2000spatially,berthier2011dynamic,sillescu1999heterogeneity,muranaka1995beta,yamamoto1998dynamics,weeks2000three,teboul2008pressure,teboul2005molecular,garrett2011three,malenkov2003structural,malenkov2002structural,jana2009intermittent,alba2003confinement,bapst2020unveiling}.
DH involves spatially separated domains of slow- and fast-moving molecules. These domains are mobile and dynamic and their correlation length-scale grows as the temperature decreases towards the glass transition temperature.
Simulations have shown that in water and other liquids there exist specific structural hallmarks that characterize immobile and mobile domains~\cite{teboul2005molecular,shi2018origin,fris2018spatiotemporal,srivatsava2019quantification}. In addition, in recent simulation studies of bulk water we have shown that the relatively immobile domains are the birthplace of ice~\cite{fitzner2019icem}.

Studies of DH have generally focused on exploring bulk homogeneous systems~\cite{giovambattista2004dynamic,giovambattista2005clusters,kawasaki2007correlation,la2004static,handle2018adam,ediger2000spatially,berthier2011dynamic,sillescu1999heterogeneity,muranaka1995beta,yamamoto1998dynamics,weeks2000three,teboul2008pressure,teboul2005molecular,garrett2011three,malenkov2003structural,malenkov2002structural,jana2009intermittent,alba2003confinement}.
Notwithstanding exciting work on highly stable vapour deposited glasses~\cite{ediger2017perspective,gutierrez2016front,leonard2010macroscopic,singh2013ultrastable,kearns2010one,zhang2016long,peter2006thickness}, the effect of interfaces on DH is less well explored.
Some pioneering studies investigated dynamical heterogeneity of water in the first adlayer at metal~\cite{limmer2013charge,limmer2013hydration,willard2013characterizing,limmer2015nanoscale} and protein~\cite{fogarty2014water} surfaces using molecular dynamics (MD) simulations.
Simulations have also shown that the diffusion of water molecules can increase in the liquid when confined in nano-porous materials~\cite{teboul2005molecular,teboul2019specific}.
Nonetheless, significant gaps in our understanding of DH at interfaces persist, despite the fact that interfaces are omnipresent and often of crucial importance to the physiochemical properties and processes of materials. For example, ice nucleation in nature happens at interfaces rather than in the bulk~\cite{sosso2016crystal,atkinson2013importance}.
In addition, the dynamics of interfacial water is of increasing importance and interest for water flow through nanometric pores and membranes~\cite{canale2019nanorheology,charlaix2017hydrodynamic,yoshida2018dripplons,fumagalli2018anomalously,neek2016commensurability,gopinadhan2019complete,zhou2018electrically,chong2018water,herr+20}, as well as at the surface of proteins and large biomolecules~\cite{qiao2019water,stanley2009heterogeneities,banerjee2020dynamical}. 
Thus, understanding how the presence of interfaces perturbs DH is of broad interest and here we seek to address the following key questions. Is DH in liquid water altered at interfaces and, if so, how does it vary from one interface to another? In addition, how is DH coupled with the structural changes that are inevitably present at the surfaces of liquids?

To answer these questions we have performed an extensive set of molecular dynamics simulations, using a classical intermolecular potential for free-standing and confined water films as well as water nano-droplets.
\pg{All the systems considered here are in equilibrium at 0\,atm and 250\,K, which corresponds to a point in the phase diagram at which the supercooling is such that DH is clearly noticeable.\footnote{This statement neglects effects of Laplace pressure which are present in the droplets. For simplicity, we use a bulk reference of 0\,atm throughout.}}
We find that interfaces impact the qualitative nature of DH, changing it from a phenomenon that spatially averages out in time to one where the differences in terms of mobility persist over time and correlate with the distance from the interface. 
In addition, we find that the extent of the observed dynamical influence of the surface depends on how the interface is defined as well as on the nature of the interface. 

\section{Materials and Methods}

The MD simulations performed in this work were done using the large-scale atomic/molecular massively parallel simulator (LAMMPS) code~\cite{plimpton2011lammps} and the TIP4P/Ice model of water~\cite{abascal2005potential}, which is especially well-suited to explore DH in supercooled liquid water given its ability to accurately reproduce the dynamical properties of even deeply supercooled liquid water~\cite{vega2005relation}.
\pg{The melting temperature of bulk TIP4P/ICE is reported as 272\,K~\cite{abascal2005potential,vega2005melting,abascal2006melting,vega2006vapor,conde2017high}; as we truncate and shift Lennard-Jones interactions at 8.5\,\AA~(see below), recent work estimates a slight increase to approximately 275\,K~\cite{atherton2022can}. For the droplets, we expect the freezing point to be depressed (see e.g., Ref.~\citenum{pan2011melting}).}
\pg{As expected, we did not observe ice nucleation in any of our simulations.}

The bulk simulation cells were first equilibrated in the NPT ensemble at 0\,atm for 10\,ns. 
We discarded the first 6\,ns for equilibration and used the last 4\,ns to compute the average volume and define the length of the cubic box used in all the subsequent NVT simulations. NVT trajectories run for a further 35\,ns, with the first 5\,ns discarded to properly equilibrate both structural and dynamical properties.
%
%

\begin{figure}[H]
\centering
\includegraphics[width=0.80\linewidth]{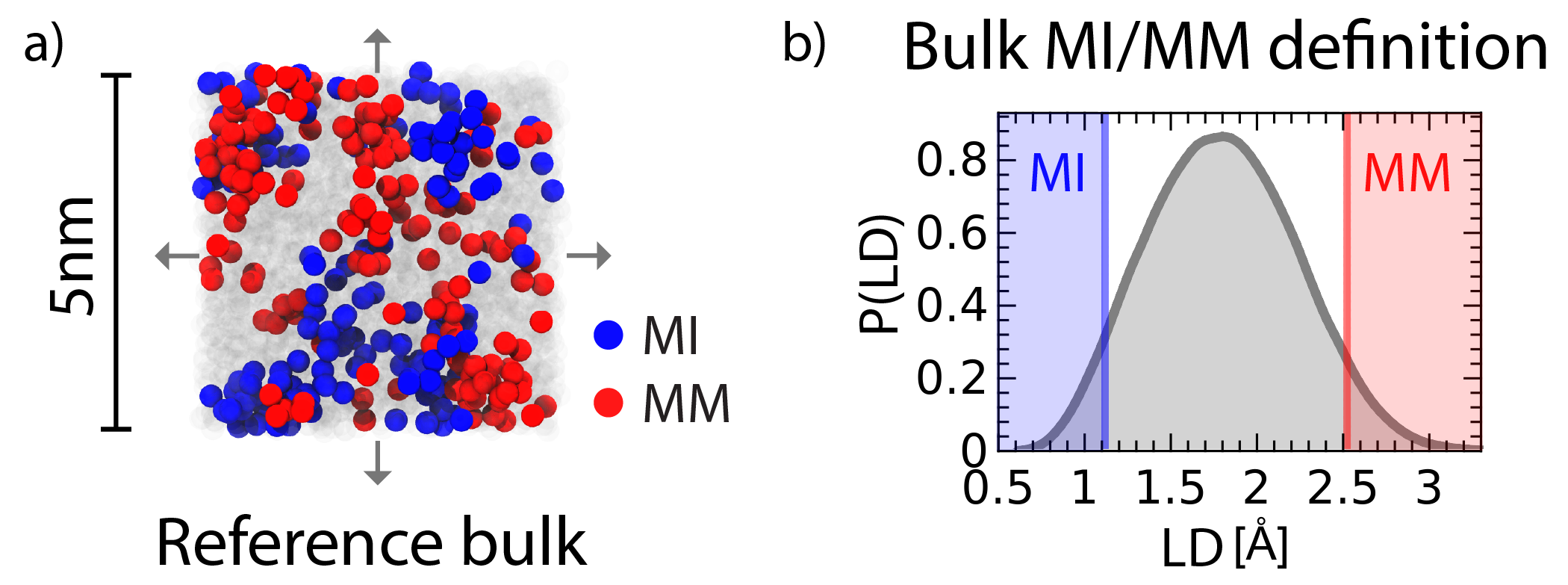}
\caption{ Dynamical heterogeneity in bulk TIP4P/ICE water. a) Snapshots of liquid water from an equilibrated NVT MD simulation at 250\,K. Only oxygens are shown and colored according to the MM (red) and MI (blue) definition shown in panel b). Gray arrows represent the directions where Periodic Boundary Conditions (PBC) are applied. b) The bulk LD probability distribution at 250\,K. Red and blue represent the top and bottom 5 percentiles, which define respectively MM and MI molecules.}
\label{fig:figonenew}
\end{figure}

\begin{figure}[H]
\centering
\includegraphics[width=0.95\linewidth]{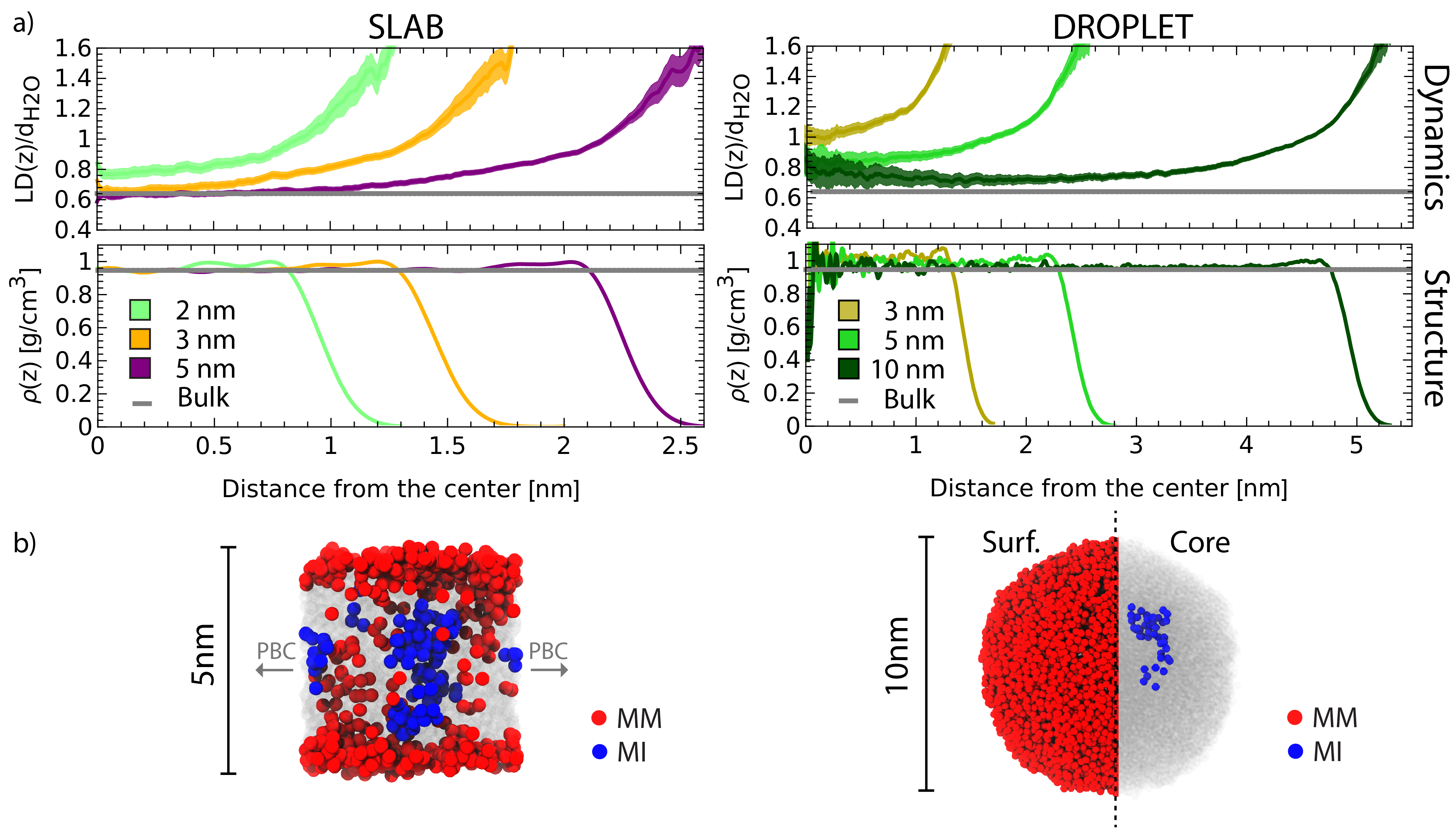}
\caption{Dynamical heterogeneity at \pg{liquid/vapor} interfaces for slab (left) and droplet (right) geometries at 250\,K. a) Top panels: The average LD profile (median value) normalised by the diameter of a water molecule (2.8~\AA) as a function of the distance from the centre for the different slabs (left) and droplets (right). Only for the 3 and 4 nm thick slabs does the dynamics away from the interface converge to the bulk value. Bottom panels: convergence of the average density profile as a function of the distance from the slab's (left) and the droplet's (right) centre. The structure converges to bulk values within 1\,nm from the interface for all the systems but the 3\,nm droplet. b) Snapshots of liquid water from an equilibrated NVT MD simulation at 250\,K. On the left is a free-standing 5\,nm-thick slab, while on the right is a 10\,nm-droplet in vacuum.  Only oxygens are shown and colored according to the MM and MI definition shown Fig.~\ref{fig:figonenew}b. Gray arrows represent the directions where PBC are applied. One can notice that DH is heavily influenced by the presence of the interfaces.}
\label{fig:figtwonew}
\end{figure}

The rigidity constraint of molecules in the TIP4P/Ice model is imposed with the Rattle algorithm~\cite{andersen1983rattle} and long-range electrostatic interactions are computed by using the Particle-Particle-Particle-Mesh (PPPM) algorithm~\cite{hockney1988computer} with non-electrostatic interactions are truncated and shifted at 8.5\,\AA; 
this setup ensures that the density in bulk simulations without interfaces (e.g., Fig.~\ref{fig:figonenew}) is equivalent to the density found in the center of the planar slabs (e.g., Fig.~\ref{fig:figtwonew}b).”

For a cubic box containing 3,072 molecules we find an equilibrium box length of 45.960\,\AA~at 250\,K and 0\,atm. 
In order to check for possible finite size effects due to the box size, we tested different lateral lengths and aspect ratios, finding that the cubic box used is large enough to converge the bulk LD distribution (see Section S6 of the SI).

The equations of motion were integrated using a 2\,fs time step and a 10-fold Nos\'e-Hoover chain with a relaxation time of 2\,ps to control temperature (see Section S7 of the SI for an in depth analysis of the effect of the thermostat on DH). For all film calculations, initial configurations are prepared from the equilibrated bulk, by increasing the \emph{c} lattice vector to add a vacuum region five time larger than the confinement length.
In the slab geometry, the Yeh-Berkowitz slab correction~\cite{yeh1999ewald} was applied, as implemented in LAMMPS.
Each slab has been further equilibrated for 35\,ns.
More information about the equilibration of the slabs can be found in the SI.
For the free-standing slab, several thicknesses have been explored in this study: 4,020 (2\,nm-thick), 5,994 (3\,nm-thick) and 9,216 (5\,nm-thick) molecules.
The same procedure has been used to equilibrate a 10\,nm-thick slab (18,432 atoms) interacting on one side with an hard LJ substrate. Indeed, the cell is long enough in the z-dimension so that the water condenses and forms both a water-substrate and a water-vacuum interface. Although the overall simulation cell is held at constant volume, the presence of the free water-vacuum interface acts as a natural barostat to the liquid.
We choose to use a 9-3 LJ potential to mimic a smooth planar solid/water interface, inspired by the recent work of Brandenburg et al.~\cite{brandenburg2019physisorption}, where they showed how water at graphitic interfaces can be modelled effectively by combining classical water force fields with LJ parameters tuned to reproduce the desired oxygen-surface interaction. Using different LJ potential walls with varying adsorption strengths, they were able to demonstrate how modest changes in the adsorption energy lead to drastic changes in the wetting properties of the surface.
Here, the standard 9-3 wall implemented in LAMMPS was used with $\sigma=3.37265$\,\AA~ and
$\epsilon=$0.1 and 0.7\,eV, thus changing the hydrophilicty of the surface over a broad range.
The LJ structured slab was built using the \texttt{diamond100} function of the ASE library~\cite{larsen2017atomic} by setting \texttt{size=(12,12,10)} and scaling the final coordinates to match the box size used for all the other planar systems at 250\,K (i.e. lateral size of 45.96\,\AA). 
The atomic positions in the substrate are kept constant during the MD trajectory.
Droplets were created from the bulk at equilibrium density (at 0\,atm) adding a vacuum region five times larger than the droplet diameter and equilibrated for ~1\,ns. 
Three droplet sizes have been explored: 1,344 (3\,nm), 6,213 (5\,nm) and 49,608 (10\,nm) atoms.

Finally, to aid the comparison, the dynamical profiles are normalized by the diameter of a water molecule (which we indicatively take to be 2.8~\AA). 

\subsection{Characterisation of the local liquid dynamics}

To compute DH directly from atomistic simulations we used the so-called iso-configurational analysis (ISOCA). 
ISOCA is a computationally demanding technique that, given a snapshot drawn from an equilibrated trajectory, allows LD to be determined for each water molecule. 
LD represents the tendency of the molecule to move and is defined as
$LD_i=\langle \|\mathbf{r}_i(t_0)-\mathbf{r}_i(0)\|\rangle_{iso}$,
where $\mathbf{r}_i(t)$ is the position vector of molecule $i$ at time $t$. 
The time of maximum heterogeneity, $t_0$, is a function of the temperature $T$ (increases with decreasing $T$) and represents the time needed, starting from a specific time frame $t=0$, to observe the most heterogeneous distribution of nearest-neighbor displacements. 
The workflow to characterize the liquid dynamics together with the values of $t_0$ for different supercoolings is reported in detail in Ref.~\citenum{fitzner2019icem}. 
Note that the value of 5\% as the threshold for MM and MI molecules does not affect the results presented in this work, as shown in SI. 
In this work the ISOCA for the different systems is done using the $t_0$ estimated for the liquid bulk at the corresponding $T$; 
this ensures consistency in the treatment of the interfacial and bulk systems.

The notation $\langle ... \rangle_{iso}$ indicates an ensemble average, i.e. we run many trajectories starting from the same initial configurations, but with different initial velocities and we eventually average the local displacements at time $t_0$ over the ensemble. Normalising the displacement of each molecule by the oxygen's mean-square displacement (MSD) one would get a similar quantity, referred to as dynamical propensity (DP). We chose not to normalise by the MSD in order to get an estimate of LD comparable between systems types.
Given the size of the boxes used in this work we carefully checked the minimum number of replicas needed in the ISOCA to converge the LD distribution in bulk. We found that at least 30 ISOCA runs are needed to obtain a converged LD (see Section S2 of the SI), with at least 10 independent starting frames necessary to converge the total LD distribution. All the results shown here where computed from 45 ISOCA. To obtain a smooth estimate of the LD($z$) profile improved statistics is needed, and at least about 30 independent frames are necessary. More information on the convergence of the LD statistics can be found in the SI. 
The value of $t_0$ varies as a function of the temperature and pressure. Following the procedure described in Ref.~\citenum{fitzner2019icem}, we find that $t_0=53$\,ps for TIP4P/Ice at 250\,K and 0\,atm.

\subsection{Definition of the intrinsic frame of reference}

To generate the intrinsic frame of reference we utilize the construction introduced in Ref.~\citenum{willard2010instantaneous} and implemented in a Python-based tool to calculate instantaneous interfaces and concentration/orientation profiles from molecular simulation trajectories in slab geometry~\cite{willard-chandler}. 
The tool uses the Lewiner marching cubes algorithm~\cite{lewiner2003efficient} and is partly an adaptation of the Willard-Chandler module of the Pytim code~\cite{sega2018pytim}.
The Willard-Chandler procedure consists of associating a continuous Gaussian density function with the discrete position of each water molecule in the system, thus obtaining a density as the sum over all the Gaussian functions. For each system's snapshot we define the instantaneous interface as the set of points on the density field whose value is equal to half the average equilibrium density of the bulk liquid at the corresponding pressure and temperature.
To measure the liquid's molecular properties as a function of the distance from the intrinsic interface we project the mean property of interest along the axis perpendicular to the instantaneous intrinsic surface.

\section{Results}

\begin{figure}[H]
\centering
\includegraphics[width=0.6\linewidth]{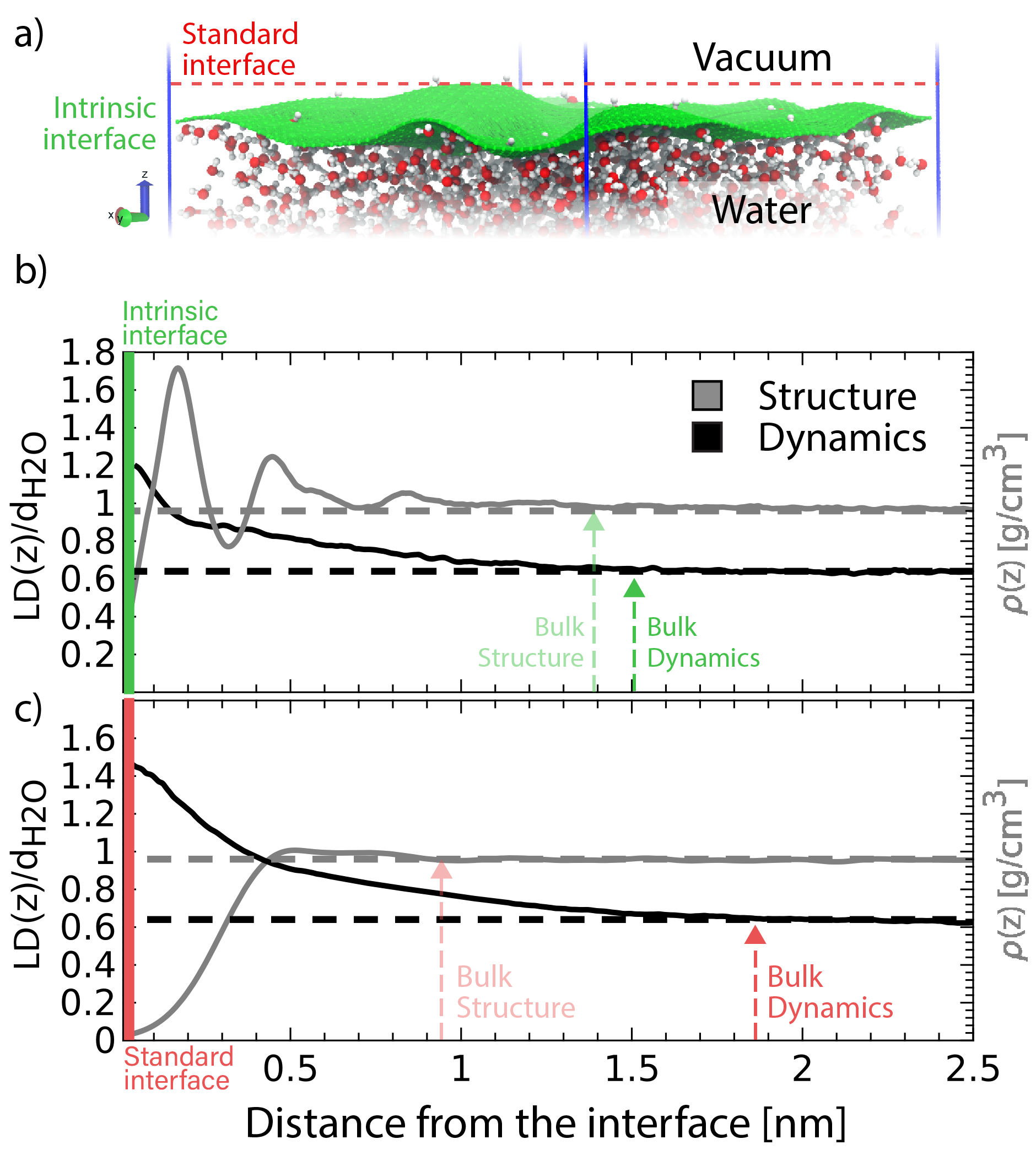}
\caption{Effect of the interface definition on both structure and dynamics for TIP4P/Ice water supercooled at 250\,K. a) Snapshot of a 5\,nm-thick free-standing slab with the instantaneous intrinsic interface rendered as a green mesh and the standard interface depicted as a red dashed line. The slab is periodically replicated in the x and y directions. b) LD profile (median value, normalized by the diameter of a water molecule) as a function of the distance from the instantaneous WC interface plotted on top of the density profile. Dashed line represent the converged values at the slab's centre. The green arrows qualitatively represent the distance at which both structure and the dynamics converge to bulk values. c) Difference between the structural and dynamical decays in proximity of the standard interface. Clearly, the dynamics appears to converge to bulk values differently from structure when using the intrinsic surface as reference.}
\label{fig:figtwo}
\end{figure}

We begin by reviewing briefly DH in bulk water. 
Fig.~\ref{fig:figonenew}a shows a visual representation of DH in bulk TIP4P/Ice water 
at T=250\,K and P=0\,atm, taken from a molecular dynamics simulation in a 5\,nm$^3$ box. 
The TIP4P/ICE model is well-suited for this study as it reproduces the melting point of hexagonal ice (Ih)~\cite{conde2017high}, as well as both the densities of water and ice and the coexistence curves~\cite{abascal2005potential}.
Following previous studies \cite{fitzner2019icem, widmer2004reproducible,widmer2005relationship}, we label each molecule as \emph{Most Immobile} (MI) or \emph{Most Mobile} (MM) when its \emph{local diffusivity} (LD) respectively falls into the lower or higher 5\% tail of the total LD probability distribution function of the bulk (Fig.~\ref{fig:figonenew}b).
As noted in the Methods section, LD is an explicit measure of the extent to which the dynamics of the liquid, over the time scale of structural relaxation~\cite{widmer2007study,fris2007short,kim2014dynamic}, is determined by the initial configuration. 
Fig.~\ref{fig:figonenew}a provides a vivid picture of DH in bulk water, which appears as distinct domains of relatively immobile (blue) and relatively mobile (red) water molecules. 
Dynamical domains, in bulk, average out in space over time, 
continuously growing, shrinking, and fluctuating in size and shape. 
The stronger the supercooling, the larger the size of the domains \cite{fitzner2019icem}, and at the level of supercooling shown in Fig.~\ref{fig:figonenew}a a single dynamical domain typically does not exceed a few nanometers, which we take to be indicative of the dynamical correlation length in our system.
This length scale should not be confused with the correlation lengths typically reported to quantify the cooperative motion of water molecules within DH clusters~\cite{berendsen1987missing,giovambattista2005clusters,scala2000configurational,starr2003recent,giovambattista2003connection,giovambattista2004dynamic} (see Section S4 of the Supporting Information, SI).
Through a careful series of studies on systems in different sized simulation boxes, we established that in the temperature range 230--270\,K, a cubic box with lateral size of 5\,nm is sufficient to converge the average dynamical properties in the bulk (see Section S1, S3 and S6 of SI).
Length scales corresponding to the size of the dynamical domains observed in bulk are, however, comparable to confinement lengths often observed in biological and technological systems ($\gtrsim$1--2\,nm)~\cite{banerjee2020dynamical,mukherjee2019mechanism,bocquet2020nanofluidics,wuttke08,ehm2021intrinsically,ball08,gao2021anomalous,ribo11}. 

In order to understand dynamical heterogeneity in supercooled interfacial water we performed a range of simulations on different types of interfaces.
Specifically we considered free-standing slabs of thickness between 2 to 5\,nm and spherical droplets of diameter between 3 to 10\,nm. Confined water films were also considered and these are discussed further below. 
A summary of the key results obtained for water films and water nanodroplets is presented in Fig.~\ref{fig:figtwonew}. 
We find that, irrespective of whether we have a nanodroplet or a free-standing film, a clear dependence of DH on the distance from the interface is observed.
This is shown in Fig.~\ref{fig:figtwonew}a (top panel) where we present the median LD profiles as a function of the distance, $z$, from the slab (or droplet) center. 
From this it can be seen that relatively mobile water molecules are found at the water-vacuum interfaces while the mobility drops upon moving into the interior of the slab or nanodroplet. 
Greater mobility of water molecules is expected at the interface with vacuum because the dynamics of breaking and forming hydrogen bonds (HBs) is faster at the water-vacuum interface than in bulk water (see e.g. Refs.~\citenum{liu2005hydrogen,abascal2005general}).
However, what we learn from the current analysis is: (i) how the recovery of bulk dynamical properties depends precisely on the size of the water film or droplet; and (ii) that the spatial arrangement of the MM and MI domains is very different in the interfacial systems (Fig.~\ref{fig:figtwonew}b) compared to the bulk (Fig.~\ref{fig:figonenew}a). 
Fig.~\ref{fig:figtwonew}a also illustrates how the mean density of the films and droplets varies as a function of $z$.
By comparing the median dynamical (top panel) and mean structural (bottom panel) profiles it appears that structural and dynamical properties converge to bulk values (gray line) on different length scales (see Section S5 of SI). 
However, this result depends on how the interface is defined.

In Fig.~\ref{fig:figtwonew}, the \pg{water/vapour} interface was defined simply by the mean density profile within a Cartesian frame of reference. We refer to an interface such as this as the ``standard'' interface.
An alternative and often more informative definition of the interface is the so-called  ``instantaneous'' interface first introduced by Willard and Chandler~\cite{willard2010instantaneous,willard2014molecular}. 
The instantaneous surface can be used to set a dynamic frame of reference, performing a spatial transformation that defines the $z$-position of each water molecule relative to the local instantaneous surface rather than a fixed Cartesian plane.
Fig.~\ref{fig:figtwo}a shows a rendering of the 5\,nm-thick slab and the corresponding standard (red) and intrinsic (green) interface. Panel b and c reveal how crucial the definition of the interface is: considering the standard interface as reference, the dynamics converges to bulk on a length scale that is twice that of structure. 
If instead, the intrinsic interface is used as reference, dynamical and structural fluctuations converge to bulk values on a much more similar length scale, with the \pg{liquid/vapor} structural profile resembling that of the liquid/hydrophobic-substrate interface.
This sensitivity of the  density profile to the definition of the interface is similar to that reported in Ref.~\citenum{willard2014molecular}, and shows that the instantaneous surface of the \pg{liquid/vapor} interface is a soft collective variable that fluctuates spatially. These flutuations induce a microscopic structure that is essentially indistinguishable from that of the liquid water interface adjacent to an extended non-polar hydrophobic substrate.~\cite{willard2014molecular}.

\begin{figure}[H]
\centering
\includegraphics[width=0.6\linewidth]{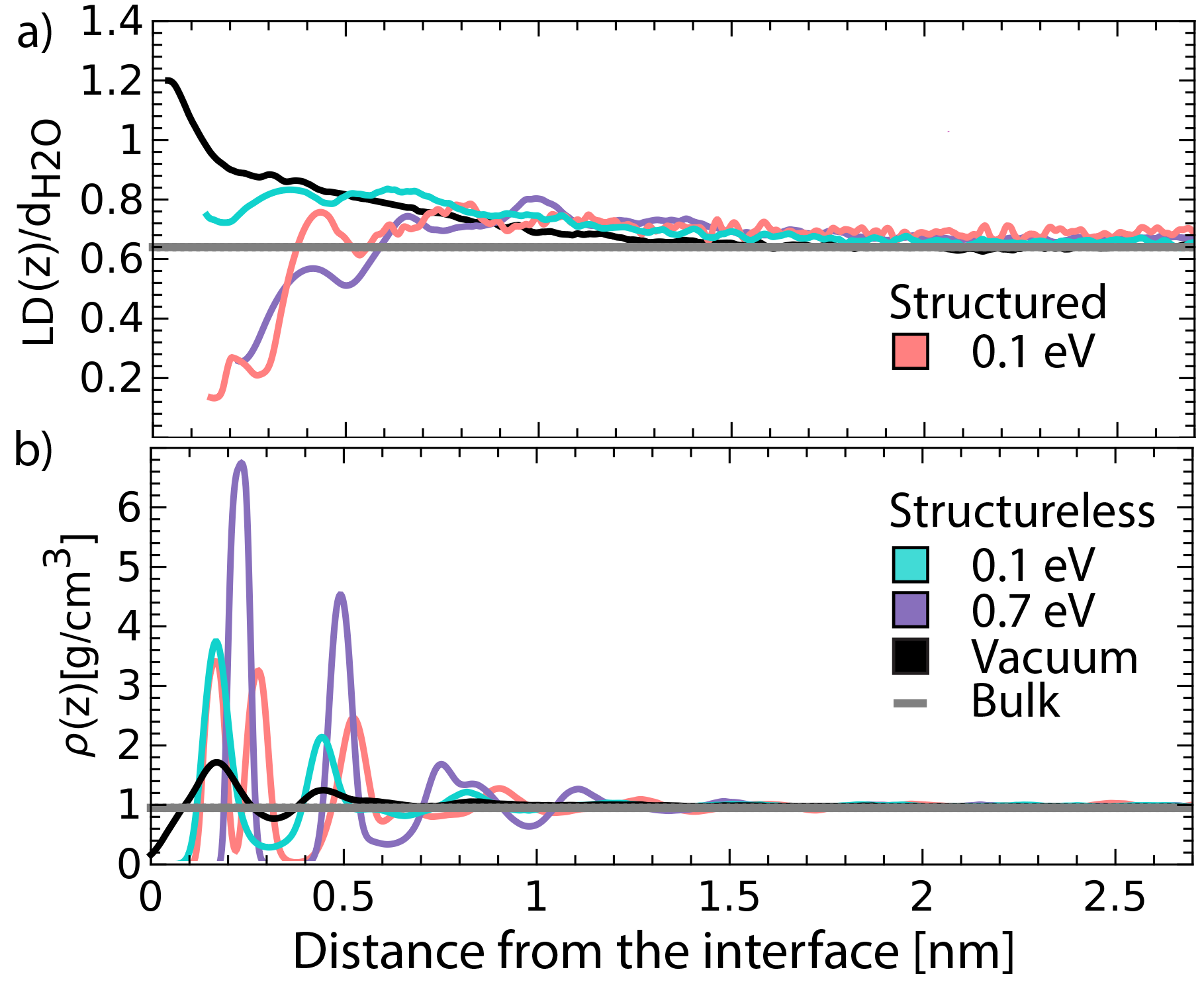}
\caption{Comparison of median LD (a) and mean density (b) profiles as a function of the distance from the slabs' centres for different interface types at 250\,K and 0\,atm. The LD value is normalised by water's molecular diameter. A free-standing (black line) 5\,nm-thick liquid slab in vacuum is compared with two structureless wall having different interaction strengths (cyan and purple line) and a structured 100 LJ/water interface.}
\label{fig:figthree}
\end{figure}

So far, we have looked at the dynamics of water in the proximity of flat or curved water-vacuum interfaces. We now discuss water films confined between solid surfaces and explore how the dynamics depends on the interaction strength of water with the interface as well as the structure of the solid substrate. 
To this end, we performed simulations of a 5\,nm-thick film of water confined between either smooth structureless walls or confined between a structured wall.
The key results of these simulations are reported in Fig.~\ref{fig:figthree}. 
Focusing on the structureless wall first, we find that: (i) there are strong density oscillations up to about 1.5\,nm from the interface; and (ii) the strength of these oscillations grows with the interaction strength.
Density oscillations such as these are well-known and have been observed many times before in both experiment and simulation studies (see e.g. Refs~\citenum{hendrik2016,lee1984structure,poynor2006water,ball08,cheng2001molecular}).
Interestingly, we find that the dynamical response of water to the substrates is quite different to the structural response:  variations in LD are much less pronounced whilst at the same time extending to slightly larger distances from the surface.  
In particular, the layering clearly seen in the density is much less apparent in the DH. 
In addition, depending on the strength of the interaction, the mobility of the interfacial water molecules can either be reduced or enhanced compared to the bulk ($cf.$ the 0.1 versus 0.7 eV data). 
Going further, in an attempt to gain some initial understanding of the difference between a structureless wall and a structured surface, we performed a simulation of water in contact with an ideal (100) surface of atoms. 
Interestingly, we find that (for the same 0.1 eV interaction strength) very different dynamical behaviour is observed: in the presence of a structured surface the mobility of the first layer drops toward zero, while at the structureless wall dynamical fluctuations are still larger on average than in bulk. 

To summarize our findings, we have seen that the presence of an interface has a profound effect on the liquid's mobility.
Our simulations of water confined between different types of interfaces suggest that there is a non-trivial link between average structural and dynamical inhomogeneities as a function of the distance from the interface.
This is governed by the strength of the interaction between water molecules, as well as by the geometry and the nature of the confining substrate. 

\section{Conclusions}

In this work, we have examined how dynamical properties of supercooled interfacial water differ from that of bulk water.
We now proceed to discuss some of the implications of these findings with a particular emphasis on ice formation and vitrification of supercooled liquid water.

Recently, Fitzner et al.~\cite{fitzner2019icem} reported that in supercooled bulk water there is a strong preference for nucleation to occur in relatively immobile regions.
The question as to where nucleation happens in interfacial water naturally arises.
Indeed, Haji-Akbari and Debenedetti~\cite{haji2017computational} investigated this specific problem using the same TIP4P/Ice water model and a 5\,nm free-standing film (in this study we have investigated the same system, see Fig.~\ref{fig:figtwonew}'s panels b and c).
Through a highly computationally demanding set of enhanced sampling simulations, they found that nucleation starts in the interior of the water film, which is in line with our observation that MI domains are found almost exclusively in the interior of the film. 

Let us now move beyond nucleation in pure water to heterogeneous systems, where a substrate or impurity in the liquid enhances the nucleation rate. Traditionally, enhancements in nucleation rates by substrates have been attributed to the structural (often templating) influence of the substrate on the liquid~\cite{lupi2016pre,lupi2014does,sosso2016crystal,fitzner2015many,cox2015molecular,bi2016heterogeneous,davies2021routes}.
Our findings provide an intriguing perspective, wherein the substrate alters the dynamics of the liquid in a manner that predisposes it to nucleation.
We suggest that a bottom-up design of the interface could allow engineering the local dynamical properties in the liquid already at moderate supercooling.
To be clear, we do not suggest that structural and thermodynamic properties are unimportant for heterogeneous nucleation. Rather, ``\emph{dynamical heterogeneous nucleation}'' may provide a complementary framework by which to investigate heterogeneous nucleation.  
In a similar vain, it would be interesting to investigate the potential role that solutes have on nucleation via their influence in dynamical heterogeneity. 
Even though the mobility of the water phase enters the nucleation rate in the form of a kinetic prefactor, which is usually overshadowed by the exponential term, the role of mobility is bound to be more and more important as we approach stronger supercoolings, where the kinetic prefactor can sometimes even outweigh the exponential~\cite{ediger2012perspective}.

Dynamical effects of supercooled water at interfaces are not only of paramount importance to ice nucleation, but they are also critical to the vitrification of supercooled liquid water into amorphous ice.
Vitrification represents one of the main approaches currently used to achieve the cryopreservation of biological material: cooling the aqueous phase down as quickly as possible helps to avoid the nucleation of potentially lethal ice crystals.
A substantial body of work (see e.g. Ref.~\citenum{elliott2017cryoprotectants} for a review) has been devoted to investigating the structure and the dynamics of supercooled liquid water at the interface with some of these systems (perhaps most prominently trehalose~\cite{magno2011understanding,vilen2013nmr}).
However, knowledge of how cryoprotectants influence the dynamical heterogeneity of the surrounding liquid phase is lacking.

On the computational side, our results also show that when investigating dynamical properties at interfaces, defining the interface in the first place can be crucial, as different definitions of the dividing surface can lead to a misinterpretation of the subtle correlations emerging between structure and dynamics; this will be most important for soft interfaces, such as the liquid/vapor interface, or biological membranes in contact with aqueous solution. 

Finally, we note that ISOCA is an extremely computationally demanding analysis technique, which limits severely the system sizes and time windows tractable, even when harnessing modern HPC facilities such as those listed in the Acknowledgements section. A great benefit for the field would come from a methodological speed up of DH analysis, e.g. by approximating the LD computation using machine learning to capture the local structure/local dynamics relationship. This would allow extracting LD for a water molecule directly from the position of surrounding molecules, thus reducing the computational cost by several orders of magnitude.

In conclusion, we have shown that the dynamics of supercooled interfacial water differs from that of the bulk and depends on the nature of the interface.
Overall, rich and subtle variations have been observed, which offers great potential for tuning the dynamical response of water through variation of the interface structure or confining material.

\section*{Conflicts of interest}
There are no conflicts to declare.

\section*{Acknowledgements}
P.G., M.F., and A.M. were supported by the European Research Council (ERC) under the European Union’s Seventh Framework Program (FP/2007-2013)/ERC Grant Agreement 616121 (HeteroIce project). S.J.C is a Royal Society University Research Fellow (URF/R1/211144) at the University of Cambridge. Computational support for our work has come from the UKCP consortium (EP/F036884/1) and the UK Materials and Molecular Modelling Hub (EP/P020194/1 and EP/T022213/1).

\bibliographystyle{unsrtnat}
\bibliography{references}  






\end{document}